# Network Agile Preference-Based Prefetching for Mobile Devices

JunZe Han[†], Xiang-Yang Li[†],[‡]
[†]Department of Computer Science, Illinois Institute of Technology, and [‡]TNLIST, Tsinghua University
Email: *jhan20@iit.edu, xli@cs.iit.edu*

*Abstract*—For mobile devices, communication via cellular networks consumes more energy, and has a lower data rate than WiFi networks, and suffers an expensive limited data plan. However the WiFi network coverage range and density are smaller than those of the cellular networks. In this work, we present a behavior-aware and preference-based approach to prefetch news webpages that a user will be interested in and access, by exploiting the WiFi network connections to reduce the energy and monetary cost. In our solution, we first design an efficient preference learning algorithm based on keywords and URLs visited, which will keep track of the user's changing interests. By predicting the appearance and durations of the WiFi network connections, our prefetch approach then optimizes when to prefetch what webpages to maximize the user experience while lowing the prefetch cost. Our prefetch approach exploits the idle period of WiFi connections to reduce the tail-energy consumption. We implement our approach in iPhone. Our extensive evaluations show that our system achieves about $60\%$ hit ratio, saves about $50\%$ cellular data usage, and reduces the energy cost by $9\%$.

## I. INTRODUCTION

According to a recent study [?], mobile web browsing is growing significantly and expected to surpass the desktop web browsing by 2015.Web browsing and news reading account for a large proportion of the time and data used by smartphones. Smartphone users with a data plan spend on average 300 minutes per month browsing the web, which is comparable to mobile voice usage [?]. The global smartphone study run by Zokem shows that web browser is the single most popular data application which accounts for 54% of data application face time and 50% of data volume for smartphones [?]. Among the user's web browsing activities, reading news makes up for 68% of the time which is one of the most frequent activities [?].

Typically there are two different networking access approaches: cellular network and WiFi network. Although the cellular network is almost ubiquitous and its coverage seems not to be a problem, the limited data plan, higher price, lower data rate and more energy consumption [?], [?] make the smartphone users prefer using the WiFi network whenever possible. It is thus natural to switch to WiFi network access whenever possible. Seamless switch between networks has been proposed to improve the networking performances [?], [?]. To further reduce the energy and monetary cost while not sacrificing the user experience, in this work we present a seamless transparent solution that automatically prefetches contents and switches between networks. Techniques of prefetching contents [?], [?] and seamlessly switching networks [?], [?] have been used previously to address various challenges. To make the content prefetching work for mobile devices, a number of challenges must be addressed. In general, we need to know when to prefetch what contents such that the user experience is not deteriorated and the overall energy cost and monetary cost for data plan is reduced. Then we need to learn and predict the availability of WiFi network access and the user web-browsing preference.

To address the challenge of knowing what to prefetch, we predict what webpages the user will visit in the near future and prefetch them via WiFi network at an appropriate time. When the user indeed wants to read these webpages later, there is no need to access them via cellular network. With the advances in mobile devices, the memory is getting larger and computation capacity are becoming stronger, so it is possible for mobile devices to learn the user's preference and prefetch the web pages in advance. Most of the previous work on web prefetch is based on the URLs of web pages visited by the user in the past. Moreover, the majority techniques for web prefetch focus on short-term prefetching, where the webpages will be visited in a short time period, *e.g.*, a few seconds to a few minutes at most. The short-term prefetch technology can be divided into two categories: probability-based and clustering-based approach. In probability-based approach, the web request sequence is assumed to follow a certain probability distribution and the future request hence can be anticipated according to the pattern. For the clustering based approach, the decision is based on the assumption that the webpages close to the previous downloaded webpages are more likely to be accessed in the future.

In this paper we focus on the long-term web prefetch which prefetch the web contents much earlier (*e.g.*, half an hour to several hours) before the user browses the webpages. An motivating example for this long-term prefetch is as follows. Assume that a user Bob goes to work at 8:00 am in the morning by train and he is used to read news on his way to the office. Assume there is no WiFi network in the train and Bob has a WiFi network at home. By profiling the news types and frequently visited pages by Bob in the train, we can prefetch these dynamic contents for Bob when he is about to leave home for work. Although contents prefetched might be obsolete or be invalided in the future, we assume that the user's experience will not drop by reading the news prefetched about an hour ago.

Unfortunately, for news prefetch, URL-based approach is insufficient because each piece of news is often assigned a unique URL and a user seldom reads the same news twice. It is hard to predict what kind of news a user will be interested in purely based on the URL of the news. In this work, we present a prefetching approach that combines the keyword-based and URL-based approach. Based on the list of most visited websites and sections by the users (*e.g.*, cnn.com, cnn.com/World), we assign different weights to different websites and sections. We also capture the user's dynamic preference on news by assigning decayed weights to the list of mostly appeared keywords in the news. We then sort the list of possible news based on a combination of the websites' weights and keywords weights. This will give us a list of news to prefetch.

The second challenge is to decide when to prefetch these contents. Several challenges make the decision of prefetch timing difficult: (1) if a user has a WiFi network access now, we need to predict when the user will move outside the coverage region of the WiFi networks; (2) even if we know when the WiFi network connection will be lost, we need to estimate the time needed for prefetching the contents so that the prefetched contents are mostly up-to-date when the user reads it; (3) the prefetching activities should not interfere the regular network usage activities by the user (otherwise, the user experience will be deteriorated).

In summary the main contributions of this work are as follows. We combine the keyword-based and URL-based prefetching approach to predict the webpage a user will be interested in. We also present a network condition prediction approach that can start contents prefetch automatically. In addition, we propose a prefetch scheduling algorithm to exploit idle time of the network to prefetch the web pages. See Section IV for details of our system design and approaches. We implement the prototype system in iPhone and conduct extensive evaluations on the performance of our system. The experimental results reported in Section V show that our system has high hit ratio and low waste ratio while at the same time is energy efficiency and data saving.

## II. RELATED WORK

### A. Desktop Prefetch

Web prefetching has been studied for a long time. The primary objective of prefetching technique in desktop computer is to reduce the latency of web page loading and increase the hit ratio of the prefetched web pages.

*1) Short-Term Prefetching:* Short-term prefetching technique aims at predicting the web pages to be visited in the near future say the next one or two web pages. Markov model is an important approach for short-term prefetching. The Markov-based prefetching technique predicts the next web pages to be visited by analyzing the past visit sequence. Xing [?] proposed a hybrid-order tree-like Markov model that can predict web access precisely. By analyzing the user's visit path, Jin [?] proposed a prefetching model based on the Hidden Markov Model (HMM) that is able to capture and mine the user's information requirement for the future and then makes semantic-based prefetching decisions.

Firefox [?] introduced a new strategy for website optimization: link prefetching. The browser can prefetch specified web pages based on the prefetching hints provided by the web page. It utilizes browser idle time to download or prefetch documents that the user might visit in the near future. In HTML5 [?] prefetching is done via the LINK tag, specifying "prefetch" as the *rel* and the *href* being the path to the document.

*2) Long-Term Prefetching:* In contrast to the short-term prefetching, long-term prefetching uses long-term steady state object access rates and update frequencies to make prefetch decisions. Markatos [?] proposed a top 10 approach for prefetching: the server keeps the most popular documents requested by the proxies and the clients, and then these top 10 documents will be considered for prefetching by the clients. In the approach by Shin [?], the proxy calculates both the most popular domains and most popular documents in those domains, then prepares a rank list for prefetching. Keyword-based prefetching for Internet news services is presented in [?] [?]. The keyword-based approach is able to prefetch the URLs that is rarely accessed or have never been before. Venkataramani *et al.* [?] presented and evaluated a long term prefetching policy based on both object request rates and lifetimes. Wu [?] proposed a H/B-Greedy prefetching to improve the H/B metric which combines the effect of increasing hit rate (H) and reducing the extra bandwidth (B) consumed by prefetching.

### B. Mobile Prefetch

The prefetching technique for mobile device also needs to be energy and data efficient. Balasubramanian [?] presented a measurement study of the energy consumption characteristics of three widespread mobile networking technologies and finds that 3G and GSM incur a high tail energy overhead incurred by staying in high power states after completing a transfer. Liu *et al.* [?] proposed the TailTheft to schedule a number of transmissions to the Tail Time of other transmissions. Balasubramanian [?] explored the use of intermittently available WiFi to reduce 3G data usage and ease pressure on cellular networks. [?] proposed the VAP scheme which can dynamically adjust the number of prefetch jobs based on the current energy level to prolong the system running time. [?] described a prefetch scheme that adapts to different network systems. [?] presented a cost-benefit analysis to decide when to prefetch based on the performance such as latency reduction, the cost of energy and monetary cost or data usage.

Lymberopoulos [?] presented a prefetching technique based on machine learning approach for mobile phone. Their prefetch technique is based on the visited URLs and only focus on the URLs visited most frequently. They extract spatiotemporal feature form the user's browsing behavior and use these features for training. Their approach also decreases the energy consumption by offloading the machine leaning job to the remote server. Song [?] proposed novel algorithms to mine the association rules and use them to construct the prefetch sets and proposed a cache-miss-initiated prefetch (CMIP) scheme to reduce system resources consumption such as bandwidth and power. Armstrong [?] employs a pair of proxies, located on the mobile client and on a fully-connected edge server respectively. The client specifies her interest by highlighting portions of already fetched webpages through the browser. If relevant changes have occurred, the edge proxy will aggregate the updates as one batch to be sent to the client. Komninos [**?**] presents a prefetching approach based on the contextual information regarding the user's activities and interests extracting from their electronic calendar.

To enhance the prefetching performance, it is also important to predict the network environment [**?**]. Higgins [**?**] presented a predictive frame for mobility-aware prefetching. It uses the Gauss-Markov model to estimate the movement and further estimate the probability of the future location.

## III. PROBLEM FORMULATION AND CHALLENGES

### A. Problem Formulation

Given the user's access history and the past network condition, we want to prefetch the web pages via WiFi network for the user to browse when the WiFi network is not available. Here we focus on prefetching news web page from news websites. To capture the user behaviors and for easy manipulation, we designed and implemented a web browser for iOS. We assume that the user visits web pages through our browser and we can keep track of the user's access history. In addition, to keep the prefetching activities transparent from the end-users while not interfering the regular browsing activities of end-users, our system should prefetch the webpages automatically by carefully exploiting the idle network connections. Our system also will predict the network connections, especially, when the end-user will have a WiFi network access and when the end-user will leave the current WiFi network. Based on this network prediction and the user browsing behavior and preference model, we should carefully schedule the prefetching jobs such that we can prefetch the most latest related news before the user leaves the current WiFi network coverage. Observe that a user still needs to use cellular network connections when the user wants to read some news webpages not yet prefetched. An intermediate goal here is to reduce such cellular network connections as much as possible. The ultimate goal is to carefully decide what contents to be prefetched at what time such that the overall energy consumption and the paid network access are reduced.

### B. Challenges and Approaches

*1) News Preference Issue:* How to learn the user's preference is a challenging problem. Learning the preference by the URLs visited by the user is not accurate for news prefetching, as a user always read the newly added news which often have new URLs, which are hard to predict. Another technique for learning the user's preference is to perform statistics (*e.g.*, the keywords frequencies [**?**]) on the news webpages visited by the user. Keywords extraction has been widely used for understanding webpages [**?**], [**?**]. Unfortunately these techniques cannot be directly applied here as our goal is to reduce energy consumption and data access: If we analyze the whole news word by word, it will cost a large amount of energy and might not find the keywords in which the user are really interested. Besides, the keyword list maintaining is also challenging since the keyword list will keep growing as the time goes on and some keywords will become obsolete. We need to refresh the keywords list such that the keyword list size is bounded and it indeed actually captures the user's recent interests.

Not only the keywords, the visit paths within a news website are also important for preference learning. It will consume a large amount of time to search the news from all news websites. Even for the same website, there exits always multiple sections for different categories and searching all the sections also costs time. Thus the prefetching need to combine the keywords and the visit path of the websites, that is we need to prefetch the news including the keywords from the user's favorite websites. Instead of keep tracking all the visited URLs, we only keep track of the URLs of the news website and the URLs of the sections and subsections in the news website (see Fig. 2 for details).

*2) Performance Issue:* Performance is a major concern due to the limited bandwidth of the mobile device. Fast loading of a webpage is what the user desires, so the prefetching is supposed not to interfere the user's normal browsing. It is better that we prefetch contents when there are no regular networking activities: prefetching should be interleaved among regular data transferring. In order to exploit the period after the data transferring is completed, we need to estimate when the next regular web browsing requests will occur and how long a prefetching job will take. Observe that a user often experiences various networking speeds due to different networking environment (*e.g.*, time, location, and APs). However, our extensive evaluations (downloading more than 300 webpages) at three different locations show that the fetch times for webpages do not have a large variance. For simplicity, in this work, we will use the mean fetch time as the estimated time for prefetching any new webpage.

*3) Energy Issue:* Due to the limited lifetime of the battery in mobile devices, energy consumption is an important consideration for the mobile prefetching approach. Prefetching web pages will consume large amount of energy when using wireless connection, especially when using cellular network or the connection is poor. WiFi networks, in general, consume less energy than cellular network, so if we can always prefetch contents via WiFi network it will save the energy consumption. However, the WiFi network is not always available for some areas. Thus, we should exploit the WiFi connections whenever possible. To ensure that the prefetched news is up-to-date, we will delay the prefetch as late as possible.

Though the WiFi network is energy saving compared to the cellular network, both WiFi and cellular network waste some energy when the data transfer is completed, which is called "tail energy" [**?**], [**?**]. Since the mobile device is at high powering setting when transferring data via WiFi or cellular network, after the mobile device completes the transferring it still consumes more energy than normal setting. Though the "tail energy" is unavoidable for the mobile devices, we can also exploit the period after the transferring to prefetch the news for the user in advance, such that the tail energy will not be wasted. Thus we also need to learn the user's behavior such as how long the user stays in the current WiFi network and the time separation between two normal consecutive networking activities.

On the other hand, prefetching the useless news that the user will not read in the future also wastes a certain amount of energy. If the news is with low possibility to be read in the future, then it is better not to prefetch it, instead let the user access it via cellular network when needed. Though the more we prefetch, the higher possibility we prefetch all the contents to be accessed, but the more energy consumption. Thus we need to balance the amount of the news to prefetch and accuracy of the prefetching.

## IV. SYSTEM DESIGN

### A. System Architecture

To predict the webpages the user will visit in the future, we need to learn the user's preference based on the news read. Besides in order to effectively schedule the prefetching jobs, we also need to learn the user' s browsing behavior. Table I shows the data we need to learn about the user's preference and behavior.

TABLE I
PREFERENCE AND BEHAVIOR LEARNING

| Learning | Data |
| --- | --- |
| Browsing Behavior | Time interval between two web requests, Enter time, Leave time time of the webpages |
| Browsing Preference | Keywords in the title, visited URLs |

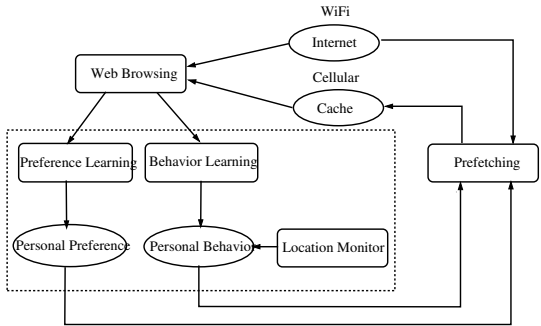

Fig. 1. System Architecture

Fig. 1 illustrates the system architecture. In our system, the preference learning module learns the user's preference and predicts what web pages the user will visit based on the preference learned. Location monitor module runs in the background to keep track of when the user enters and leaves a certain Wifi coverage area. Based on the historical log provided by the location module, the behavior learning module will predict when the user is about to leave the wifi coverage area and trigger the prefetching module to start prefetching. If the prefetching module is triggered to start, it will find the web pages that is likely to be visited in the future based on the prediction results provided by the preference learning module. After extracting the URLs of the web pages to be visited, the prefetching module then schedules these prefetching jobs according to the user's browsing behavior which is provided by the behavior learning module.

### B. Network Environment Prediction

The first challenge to be solved in our system design is decide on when to prefetch web pages for the user. Although the user wants to read the latest news, but for most users they will be also interested in the news happened several minutes or hours ago. So prefetch the "old" news is reasonable and meaningful to the user. But we still want the news prefetched to be relatively new, so we begin the prefetching as late as possible while ensuring the prefetching can be finished before the user leaves the WiFi. To do this, we need estimate how long the user will stay in the coverage area of the current network. For most of the people, the daily schedule and weekly schedule do not change a lot and hence it is possible to do such network coverage prediction.

Monitoring the network connection all the time sometimes may give us misleading information about how long the user stays in the WiFi area, because it is possible that in a certain position in the WiFi coverage area, the WiFi access is temporarily unavailable. Instead of monitoring the network connection, we monitor the user's current location to estimate when he or she leaves a certain area. The location module is available in most of the mobile devices that can help us to monitor when the user enters of leaves a certain area.

We observe that the durations of a user staying in a network depend on the time. For example, in weekdays a user may go to the office at 9:00 am and work till 12:00 pm during which the WiFi network is available. At noon the user might go outside to have lunch where he would like to read the news but the WiFi network is not available. We can then predict that the user will usually have Wifi access from 9am to about noon. In the afternoon, the user may stay in office from 1:30PM to 5:30PM. Thus, to precisely estimate how long a user will stay in a certain area, we will keep track of the following information $\langle t_i, d_i, L_i, SSID_i \rangle$, where $t_i$ is the time the user enters the coverage area of the network with $SSID_i$ at location $L_i$, and the user will stay in this network for a duration of $d_i$. We will round time $t_i$ to hours. Based on the collected networking access data, we then predict when the user will leave the current Wifi network.

As a result, for the same WiFi we have multiple time duration records for the same entering time. We use a probability-based algorithm to estimate the time the user will leave the current network, such that we can finish the prefetch in time. Let $T_i^t$ be the collection of time the user stays in the WiFi coverage area of $SSID$ $i$ after entering at time $t$, where $d$ is the day (*e.g.*, Monday, Tuesday, *etc.*) and the hour when the user enters the WiFi coverage. Let $p_i^t(d)$ be the probability that the mobile device is connected to this AP for over time $d$ when entering at time $t$. We estimate the time that the user

will stay in the current WiFi coverage as

$$\arg\min\{d \mid p_i^t(d) > \delta\},$$

where $\delta$ is a threshold value (chosen as 0.5 in our implementation).

Given the estimated staying time, it is still necessary to choose an appropriate time during this period to prefetch web pages. Prefetching late can result in that some webpages the user might browse in the future cannot be prefetched before the user leaves the WiFi coverage area. However early prefetching can also cause problems: the prefetched webpages may be out of dated as the webpages be updated; the user may browse the prefetched news when within the WiFi coverage area, which go against our purpose to prefetch webpages for the use in a non-WiFi environment. Thus we seek to find the time to start the prefetching as late as possible, while at the same time to ensure that the prefetching can be finished in time. Given the estimated staying time $d$, number of web pages to prefetch $l$ and the average fetch time $f$, in order to finish prefetching these webpages before the WiFi coverage is not available, we need to start to prefetch no later than $d - l \cdot f$.

### C. Web Page Access Prediction

*1) Preference Learning:* News title is always concise, informative and highly related to the content of the news, and the user is always guided by the keywords in the title. For example, a basketball fan may not want to miss any news titled with "NBA" or "Lakers". Thus by extracting keywords from the titles we can effectively learn the user's preference and predict what news the user will be interested in. In addition, the users are interested in different keywords to different degree and hence we quantify the keyword interest by assigning each keyword $w$ with a interest weight $q(w)$. Each time the keyword $w$ appears in the title of the news read by the user, we increase $q(w)$ by a constant $c_q$

In order to get the title of the news to extract keywords, for each requested news webpage, we will parse the head of the HTML file of the webpage and get the title of the file which is the news title. By getting rid of the function words and punctuations we can obtain the according keywords in the title.

Since people's interests will change as time goes on, newly appearing keywords in the title play an more important role in learning the user's preference than old ones. For example a user might be interested in the presidential election in November, so the news tilted with "election", "Democratic" and "Republican" will be browsed with high possibility. However in the next month, the user might be bored with the "election" and turn the interest to the "Christmas" and "new year". To deal with the issue of interest changing, we use a time decay function to keep reducing the keyword's interest weight as time goes on.

In each time period $t$, the keyword $w$'s weight $q(w)$ in the keyword list will be reduced to $q(w) \cdot (1 - \delta)$, where $1 - \delta$ is the decay rate, and remove the keyword if its interest weight is less than the threshold $\epsilon$. Not only the decay function helps us capture the user's current interests, but also contributes to reducing the size of the keyword list. To effectively maintain the keyword list, we use the heap data structure to store the keywords. In every time period we will check whether the root's interest weight is less than $\epsilon$, if so we remove it from the keyword list and heapify the keyword list, and then check the new root's interest weight until the root's weight is larger or equal to $\epsilon$, as shown in Algorithm 1.

---
**Algorithm 1** Keyword
---
1: **while** The weight of the keyword at root $q(w) \leq \epsilon$ **do**
2:    Remove keyword $w$ from the keyword list
3:    Heapify the keyword list
4: **for** each keyword $w$ in the heap **do**
5:    $q(w) \leftarrow q(w) \cdot (1 - \delta)$
---

In our keyword maintaining approach, the keyword $w$ might be removed from the keyword list and later be added again. Thus $w$'s weight $q(w)$ is accumulated and decayed since the latest time it is added to the keyword list and $q(w)$ is less than the "actual" weight. However in our approach, if keyword is removed from the keyword when weight $q(w)$ is smaller that $\epsilon$, the actual weight is also less than a constant of the removal threshold $\epsilon$. Formally, for each keyword $w$ in time interval between time $i$ and $n$ we define a weight function $f_i^n(w) = \sum_{j=i}^{n} x_j(w) \cdot (1-\delta)^{n-j}$, $x(w) \in \{0, 1\}$, where $x_j(w)$ indicate whether $w$ appears at time $t_j$. In each time slot, if $f_i^n(w) < \epsilon$, where time $i$ is the first time word $w$ appears since the last removal, the word $w$ will be removed from the keyword list. Then we have the following theorem

*Theorem 1:* If a word $w$ is removed from the keyword list at time slot $n$, then the weight $f_1^n(w) \leq \frac{\epsilon}{1-\epsilon}$.

*Proof:* Assume that a word $w$ is added into the keyword list and then removed for $n$ times. Afterwards, at time slot $a_i$ word $w$ is added into the list again and at time slot $r_i$ removed from the list. Let $v_i(w) = f_{a_{i-1}}^{r_i - 1}$ and $t_i = r_i - a_i$, then we have

$$f_1^n(w) = \sum_{i=1}^{n} v_i(w) \cdot (1-\delta)^{\sum_i^n t_i} \quad (1)$$

$$\leq \sum_{i=1}^{n} \epsilon \cdot (1-\delta)^{\sum_i^n t_i} \quad (2)$$

$$\leq \epsilon \cdot \sum_{i=1}^{n} (1-\delta)^{\sum_i^n t_i} \quad (3)$$

$$= \frac{\epsilon}{1 - \epsilon} \quad (4)$$

This finishes the proof. ∎

Since word $w$ is added at time $a_i$ and removed at time $r_i$, $(1-\delta)^{t_i}$ must be less than $\epsilon$ and hence $\sum_{i=1}^{n}(1-\delta)^{\sum_i^n t_i} \leq \frac{1}{1-\epsilon}$.

The user's preference on news is also related to sections of the news website. The news website is always organized as a tree structure as shown in Figure 2, which consists of several sections such as Sports, Economy, Entertainment, and each section also has some subsections . Learning the user's preference only based on keywords is no enough. For example the user always visits the sports section and thus the news in this section will be browsed by the user with higher probability. Thus we also keep track of URLs of the sections and subsections visited frequently by the user such that we can prefetch the news from these URLs. Similarly to the keywords, we assign a interest weight $t(s)$ to each section $s$. We keep track of the times the section $s$ is visited by the user and each time the user visits the section $s$, we increase the weight $t(s)$ by a constant $c_t$. Besides, the weights of URLs are also decayed as time goes on using the same approach for the keyword.

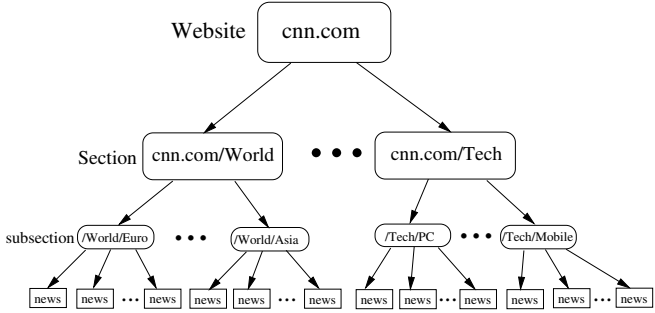

Fig. 2. Website Architecture

With the interest weights of keywords and URLs, we define the interest weight for each news according to the keywords appearing in the headline and the URL where the news appear. The higher the weight is the more possible the user will read it. Assume that keywords $w_1, ..., w_n$ appear in the news $h$'s title, and the weight of the section where the news appears is $t(s)$, then we set $h$'s weight as

$$w(h) = \sum_{i=1}^{n} q(w_i) + t(s) \quad (5)$$

*2) News Searching:* If the prefetching module is about to start prefetching, it will search the news from the sections frequently visited by the user. In order to search the news from a web page, we first need to parse the HTML files of the according sections' webpages. The HTML file is structured as a tree structure – dom tree and each element of the HTML file is represented as a node, such as title, data, image, hyperlink, etc. We search the whole dom tree to find all the nodes tagged with the $<a>$ which represents the hyperlink node. Then from the hyperlink nodes we can obtain the URL of the news webpage and the title of the news. One problem concerning the extracted URLs is that for some websites, the URL of the hyperlink might be the relatively path not the absolutely path of the webpage and prefetching the relative URL can lead to web request failure. To deal with this problem, before adding a URL to prefetch queue we need to check whether the URL is a relative URL. As the relatively path does not have the prefix of "http://", so for the relative path URL we append it to the URL of its parent node to get the full URL.

Given the news searching results, we need to decide which news webpages to prefetch. Obviously, the more we prefetch, the more likely that we can prefetch the news the user will read in the future, but the more energy cost and storage cost. To make prefetching decision, we compare the news's weight with a prefetch threshold, if it is greater than the threshold, we then add the URL of the news web page to the prefetch queue. In order to set the value of the prefetch threshold appropriately, we keep track of the weight of each news the users read to learn what weight the user is likely to read and set the prefetch threshold based on these weights. Let $p$ be the expected probability that the prefetched webpages will be accessed, $n$ be the total number of webpages the user visited and $v(w)$ be the number of webpages whose weight is above $w$, then we set the threshold weight $s$ as

$$\arg\min\{w \mid v(w) > p \cdot n\}.$$

*3) Scheduling Prefetching:* Given the prediction result from preference learning module and behavior learning module, we know the time when the user will leave the current WiFi coverage area and the collection of URLs to prefetch. Due to the limited bandwidth of the mobile devices, prefetching should not interfere user's normal web browsing as aforementioned. Observing that the network connection is idle between two web requests, we exploit the interval between two web requests to prefetch web pages, as shown in Figure3.

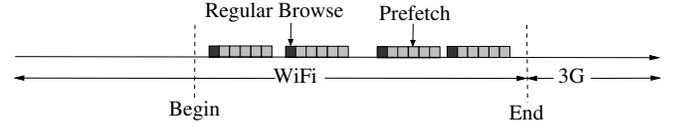

Fig. 3. Prefetch Scheduling

When scheduling prefetching jobs, we divide the time domain into time-slots of fixed length, and assume that fetching one web page costs one time-slot . Formally, let $a(u) = [a_0, a_1, ..., a_n] \in {0, 1}^n$ be the webpage access sequence by the user, and $p(u) = [p_0, p_1, ..., p_m] \in {0, 1}^n$ be the prefetch sequence. In order to not interfere the user' s normal browsing, for each time $t$, we should have

$$a_t + p_t \leq 1 \quad (6)$$

To fully exploit the idle time when the user is browsing news, we need to maximize

$$\sum_{i=1}^{n} a_i + \sum_{j=1}^{m} p_j \quad (7)$$

In our system, we maintain a queue of the webpages to prefetch. We first sort the webpages in prefetch queue in decreasing order of the weight such that we could first prefetch the web pages that are more likely to be accessed in case that we do not have enough time to prefetch all the webpages in the prefetching queue. Each time the webpage loading is finished and the network connection is idle, prefetching module will schedule a batch of web pages in the prefetch queue. In particular, the average time interval the user spend on one web page is $T$ and the average fetch time for prefetching a web page is $t$, then we schedule $T/t$ jobs from the news queue to prefetch after each normal webpage request from the user.

## V. System Implementation and Evaluation

We implement the prototype system in iPhone 4 with IOS 5.1.1. In order to monitor when the user enters and leaves the WiFi coverage area, we use a service called region monitoring provided by IOS [**?**]. In iOS 4.0 and later, applications can use region monitoring to be notified when the user crosses geographic boundaries. We use this capability to keep track of the time the user enters and leaves a certain WiFi area.

Our prototype system is used by three different people for two weeks. Everyday they browse news webpages using our system for about half an hour. When the user enters a WiFi coverage area for the first time, the user needs to register this area with its SSID in our system and we set the radius of the WiFi area to 100 meters. Before the system starts to prefetch webpages, it will send a notification to the user to ask for the permission. In addition we also allow the user to start prefetching manually.

### A. Browsing Behavior Learning

*1) Web Request Interval and Web Fetch Time:* We first present the user's browsing behavior statistics when using our system. Figure 6 shows the time intervals between two web requests when the user browses news webpages. For $65\%$ of the time intervals, the lengths of the intervals are less than 1.5 second. These short intervals usually appear on the way the user goes to the destination section after entering the news website, because the user only stays on intermediate webpages for a short time. These short intervals also appear when the user switches from one section to another section. For the rest $35\%$ time intervals, the average length is 61 seconds, which is the average time the user spends on reading one piece of news.

Figure 7 plots the time of fetching webpages via WiFi network. As we can see, for most of the web requests, the time of fetching a webpage via WiFi network is short. It takes less than 1 second to fetch a webpage for over $94\%$ of the web requests. This is because that news webpages designed for the mobile devices usually simply consists of words and one or two pictures. Thus the size of a webpage is small and hence it takes very little time to fetch a webpage via the high speed WiFi connection.

*2) Keyword Maintaining:* Figure 8 shows the interest weights of the keywords that have appeared in the title of news read by the user. In our experiment, we set the removal threshold to 0.2 and the decay period to 3 minutes. Here we define the compression rate as the proportion of the number of the keywords maintained in the keyword list to the total keywords that have appeared in news titles. Our experiment shows that we achieve the compression rate of $81\%$

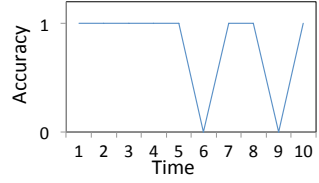
Fig. 4. Network Prediction

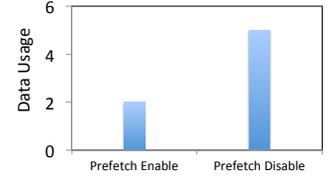
Fig. 5. Data Usage

### B. Performance

*1) Network Condition Prediction:* In our experiment, the user's daily schedule is almost fixed and hence the time the user stays in a certain WiFi coverage does no vary a lot. As mentioned above, our system will send a notification before the user's estimated leaving time and ask for the permission to start to prefetch. We assume that the network prediction is correct if the user agrees the prefetch request. Figure4 shows the accuracy of the network prediction. We achieve accuracy rate of 80% in a week for one user.

*2) Hit Ratio and Waste Ratio:* Prefetching systems are often evaluated in terms of the hit ratio and waste ratio. Hit ratio refers to the proportion of the number of prefetched webpages that are accessed by the user to the total requested webpages. Waste ratio refers to the proportion of the number of undesired prefetched webpages to the total prefetched web pages. Figure 9 plots the hit ratio of our system. As we can see the hit ratio is relatively stable and around to $60\%$ on average. Though at the beginning phase, our system does not learn the user's preference precisely and the prefetching is supposed to be not accurate. However at the beginning phase, the prefetch threshold is low due to the low weights of the news the user browse and thus the amount of prefetched webpages is large. Besides, at the beginning phase, the weight of the section is dominate when make prefetch decisions and most of the news in the user's favorite section are prefetched. Thus we can still achieve high hit ratio at the beginning phase.

Figure 10 plots the waste ratio of our system. The waste ratio here is calculated for each batch of prefetched webpages. Compared with the hit ratio, the waste ratio continuously decreases. As time goes on, the user is more likely to access to the webpages from the sections with large weight and also to access the news with titles containing keywords with large weight. Thus the prefetch threshold become larger and less webpages are prefetched. However since our system learn the user's preference more precisely, the webpages prefetched are more likely to be accessed by the subscriber and the waste keeps decreasing.

In our prototype system, we remove prefetched web pages and the clean the prefetch cache everyday. Figure11 plots the everyday's cache usage in 9 days. Similarly to the waste ratio, the cache size is relatively large at the beginning. As the system learns user's preference more precisely and the prefetch threshold become larger, less webpages are prefetched and the cache size keeps decreasing. For the last 5 days the size of the prefetch cache is about 11 MB per day.

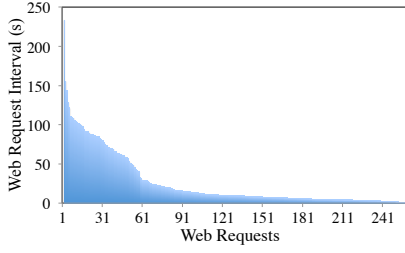

Fig. 6. Web Request Interval

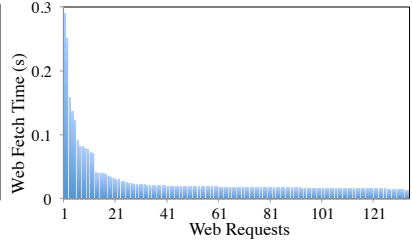

Fig. 7. Webpage Fetch Time

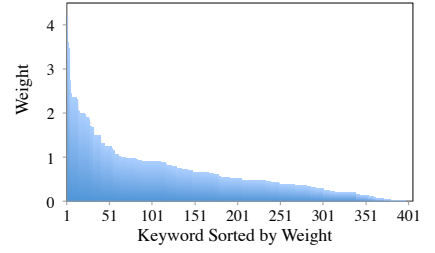

Fig. 8. Keyword Weight

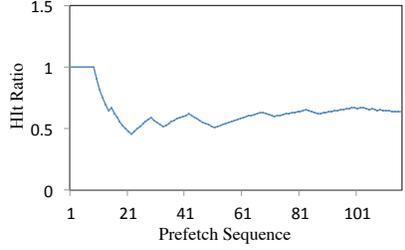

Fig. 9. Hit Ratio

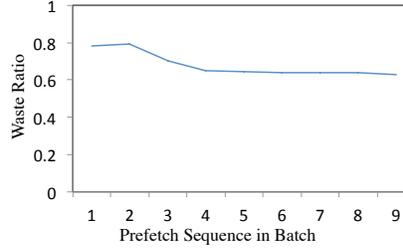

Fig. 10. Waste Ratio

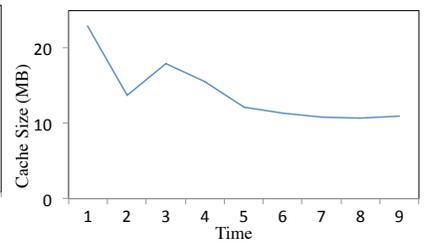

Fig. 11. Cache Size

*3) Energy and Data Consumption:* We use the *instruments* provided by the Xcode to evaluate the energy consumption. The energy consumption in iPhone is divided into 20 levels in *instruments*. In our experiment, we have two users browse the same webpages using our system with prefetch feature enabled and disabled respectively, and we close all the other applications in iPhone. Both of the users first use the system for about 10 minutes via WiFi connection and then via cellular connection for 20 minutes. When the prefetch feature is enabled, our system will prefetch webpages via WiFi network in the first 10 minutes.

Figure12(a) and Figure12(b) plot the energy consumption when the user browses webpages via WiFi network with prefetch feature enabled and disabled respectively. In both figures the peaks appear when the browser loading webpages and after loading the web page, the energy level decrease to a low level. As we can see, there are some time slots where the energy level is higher after webpage loading, this is caused by scrolling the web page which makes energy level increase due to display contents change. With prefetch feature enabled, the device stays at high energy level for longer time than the one with prefetch feature disabled. This is because our system is prefetching web pages in the background. However, as we can see that even with prefetch feature disabled, the energy level does not fall to a low energy level instantly due to the effect of "tail energy". The average energy level with prefetch feature enabled and disabled is $12.5$ and $9.6$.

Figure 13(a) and Figure13(b) show the energy consumption when our system is connected via cellular network. In Figure 13(a) after loading a webpage from prefetch cache, the energy level falls instantly to the low level. However, in Figure13(b), after fetching a webpage via cellular network, thought the energy level falls to a lower level, it still higher than that when the webpage is loaded from the prefetch cache. The average energy level when when prefetch feature is disabled is about $11.8$ while the energy level is $10.2$ when prefetch feature enabled.

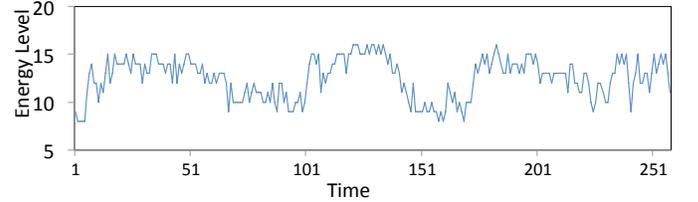

(a) Prefetch Enabled

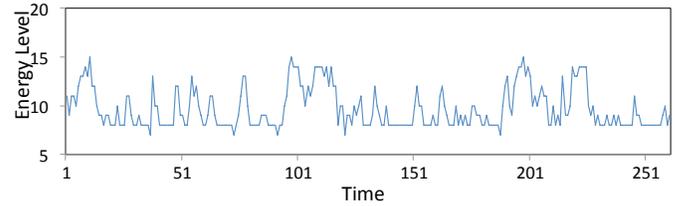

(b) Prefetch Disabled

Fig. 12. Energy Consumption via WiFi Network

Let $e(p) = e_w(p) + e_c(p)$ and $e(n) = e_w(n) + e_c(n)$ be the energy cost when the prefetch feature enabled and disabled, where $e_w$ and $e_c$ is the energy cost via WiFi connection and cellular connection respectively. We then calculate $e_p/e_n$ as the energy cost reduction. The result shows that our system consume less energy when prefetch enabled than that when disabled and we achieve about $9\%$ of the energy reduction.

During the data usage test we have two users browses about $100$ webpages in one day with prefetch feature enabled and disabled. Figure 5 plots the cellular data usage when prefetch feature is enabled and disabled respectively. When the prefetch feature is disabled, 30% of pages are not prefetched and all the other webpages are prefetched via Wifi network. The cellular data usage is 2 MB. When the prefetch feature is disabled, all of the web pages are fetched via the cellular network and thus the cellular data usage is 5 MB that is over 2 times of the one with prefetch feature enabled.

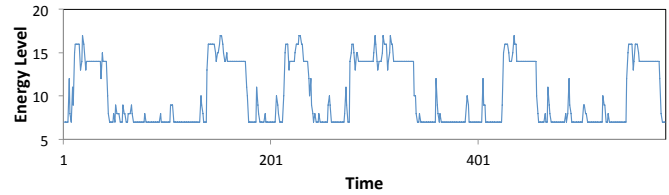

(a) Prefetch Enabled

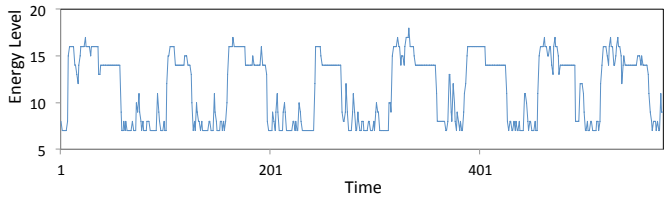

(b) Prefetch Disabled

Fig. 13. Energy Consumption via Cellular Network

## VI. CONCLUSION

In this paper we designed a network-agile preference-based prefetching method for mobile devices. We implemented our method in iPhone and conducted extensive evaluations on the performances of our methods. Our evaluations show that our prefetching based approach is able to reduce the cellular network access by about 50% and reduce the energy cost by about 9%.